\documentclass[prl,showpacs,twocolumn]{revtex4}
\newcommand{\ua}{\uparrow}
\newcommand{\da}{\downarrow}
\newcommand{\ra}{\rangle}
\newcommand{\la}{\langle}
\begin{document}
\title{
Quantum cloning of identical mixed qubits
}

\author{Heng Fan$^{\dagger }$, Baoying Liu$^{\S }$, and Kangjie Shi$^{\ddag }$}
\affiliation{
$^{\dagger }$Beijing National Laboratory for Condensed Matter Physics, Institute of Physics,
Chinese Academy of Sciences, Beijing 100080, China.\\
$^{\S }$Institute of Photonics and Photon-Technology, Northwest University, Xi'an 710069, China.\\
$^{\ddag }$Institute of Modern Physics, Northwest University, Xi'an 710069, China.
}

\begin{abstract}
Quantum cloning of two identical mixed qubits $\rho \otimes \rho $
is studied. We propose the quantum cloning transformations
not only for the triplet (symmetric) states but also for
the singlet (antisymmetric) state. We can copy these two
identical mixed qubits to $M$ ($M\ge 2$) copies. This
quantum cloning machine is optimal in the sense that
the shrinking factor between the input and the output single
qubit achieves the upper bound. The result shows that
we can copy two identical mixed qubits with the same quality as that of
two identical pure states.
\end{abstract}

 \maketitle

No-cloning theorem
is one of the most fundamental theorems in quantum mechanics and in
quantum computation and quantum information\cite{WZ}. Due to the
no-cloning theorm, it is possible for us to design quantum
cryptography such as BB84\cite{BB84}, 6-state\cite{Bruss} quantum
key distributions and various of their generalizations. The no-cloning
theorem is also
closely related with no-signaling theorem in quantum
mechanics\cite{SBG}.

In case we want to copy a quantum state, we cannot copy it perfectly
but approximately\cite{BH} or probabilistically\cite{DG}. In the
past years, much progress has already been made in designing quantum
cloning machines for different purposes[1-14], for reviews and
references, see \cite{SIGA,CF}. Bu\v{z}ek and Hillery proposed a
quantum cloning machine with one qubit input and two qubits
output\cite{BH}. The quality of the copies is independent of the
input state. This quantum cloning machine is called universal
quantum cloning machine (UQCM). Later this UQCM was proved to be
optimal\cite{BDEFMS}. For UQCM, the copies are not the same
as the input state, but this coping task can always succeed. A
different quantum cloning machine was proposed in Ref.\cite{DG}, while the coping
task can succeed with a probability, but if it succeeds, we can
always obtain perfect copies. This kind of quantum cloning machine
is called probabilistic quantum cloning machine \cite{DG}. 
Other cloning machines such as the asymmetric quantum cloning and the 
phase-covariant quantum cloning are also studied in the past years
\cite{C,BDMS,FMWW,FIMW,DP,DDZL}.
In this paper, we will restrict ourself to the UQCM case.

Bu\v{z}ek and Hillery's UQCM is for one to two case (one input qubit
and two output qubits). Gisin and Massar \cite{GM} proposed a $N$ to
$M$ ($M\ge N$) UQCM and it is also proved to be optimal by different
methods\cite{GM,BEM}. Werner\cite{W} proposed a general $N$ to $M$
UQCM not only for qubit case but also for a general quantum state in
d-dimensional system. This quantum cloning machine is realized by
symmetric projections and it is proved to be optimal for two
different fidelities\cite{W,KW}. Fan {\it et al} \cite{FMW} proposed
a $N$ to $M$ UQCM following the transformations given in
Refs.\cite{BH,GM}. This UQCM is optimal for identical pure states and
also for quantum states in symmetric subspace\cite{F}. It can be
realized by some physical systems like photon stimulated
emission\cite{SWZ,FWMI}. The super-broadcasting of mixed qubit
states which is closely related with cloning machine was recently
considered in Refs.\cite{DMP,BDMP} based on the result of Ref.\cite{CEM}. 
The experiments of
UQCM were performed in several groups \cite{LSHB,DBSS,FGRSZ,DPS}.

While considerable works have already been done to study various
quantum cloning machines, see recent review papers \cite{SIGA,CF},
there are still some simple and basic unsolved problems. The
simplest case is perhaps to copy two identical mixed qubits $\rho
\otimes \rho $ optimally. Since the UQCM proposed by
Fan {\it et al}\cite{FMW} only
provides the cloning transformations for symmetric input states, we
can copy arbitrary identical pure states and a mixed state in
symmetric subspace. If the input are two identical mixed qubits, we
cannot use this UQCM, since one input state is the singlet state
which is not in the
symmetric subspace. One may consider to simply use Werner \cite{W}
UQCM for this case and do not care about the real input, we can show
however that this method does not work. The simplest example is for
case 2 to 2 UQCM, actually we do not need to do anything and the
cloning is perfect. Here we use this example since all known UQCMs
do work for this case given the input is within their working area,
i.e., all known UQCMs can copy the input perfectly. We may find for
case $\rho \otimes \rho $, the antisymmetric states are simply
deleted by the symmetric projection operators by Werner's UQCM. This
leads to a result that the output state is different from the input
state. Thus we may find: This UQCM is not universal again for this
case, or it is not optimal. In this paper, we will consider this
problem. And we will give an optimal UQCM which can copy two
identical mixed qubits.

We should note the work in Ref.\cite{CEM} and recent results about
the superbroadcasting of mixed states in Refs.\cite{DMP,BDMP} which
are closely related with quantum cloning of mixed states.  
Those results are different from our results
in this paper. The main difference between our method and
the method in Ref.\cite{CEM} and
in Refs.\cite{DMP,BDMP} is the following: In Refs.\cite{CEM,DMP,BDMP} the input
identical mixed qubits is first divided into two groups which can
be in tensor product form by
mixed states purification. One group is the purified state need be cloned 
and another group is state which contains no information and will not be cloned.    
While the method in the present paper is that no matter whether
the input state contains information or not,  all purified states will be cloned.
And in this sense, our result is {\it universal} since the cloning procession does
not depend on the input. 
We remark that both methods in Refs.\cite{CEM,DMP,BDMP} and in the present paper are 
reasonable. They can be used for different purposes.    

{\it A 2 to 3 UQCM for mixed states.}-- A mixed state can be copied
by the same cloning transformation as we copy a pure state. Thus the
simplest non-trivial cloning task of mixed state is to copy {\it
two} identical mixed states. For this aim, we not only need the
cloning transformations for triplet states in symmetric subspace but
also need a cloning transformation for the singlet state. We
consider the UQCM in the sense that the quality of the copies is
independent of the input states. Since we consider arbitrary mixed
qubits as input, each output state $\rho ^{(out)}_{red.}$ and the
input $\rho $ should satisfy the scalar form to satisfy the
universal condition\cite{BEM},
\begin{eqnarray}
\rho ^{(out)}_{red.}=f\rho +\frac {1-f}{2}I,
\label{scalar}
\end{eqnarray}
where $f$ is the shrinking factor, $I$ is the identity. The
relationship between each input and output state is just like the
input state goes through a depolarizing channel. We can find that
the shrinking factor $f$ can describe the quality of the copies. If
$f=1$, the output state is exactly the input state. If it is zero,
the input state is completely destroyed, i.e., the output state is a
completely mixed state which contains no information. Our aim is to let
the cloning machine achieve the maximal shrinking factor. The
optimal shrinking factor has already been obtained in Ref.\cite{BEM}
for identical pure input states. It is obvious that the optimal
shrinking factor for identical pure states is also an upper bound
for identical mixed states. The problem is whether this bound can be
saturated or not for the case of  two identical mixed qubits, i.e.,
can we copy identical mixed qubits as the same quality as we copy
identical pure states?

To present our result explicitly, we first give the result for 2 to
3 cloning machine, we have 2 input states and 3 copies which may be
entangled. We consider $\rho $ to be an arbitrary mixed state
\begin{eqnarray}
\rho =z_0|\ua \rangle
\langle \uparrow |+z_1|\uparrow \rangle
\langle \downarrow |+z_2|\downarrow \rangle
\langle \uparrow |+z_3|\downarrow \rangle
\langle \da |,
\end{eqnarray}
with the restriction that this is a density operator. We also use the
notations $\chi _0=|\uparrow \uparrow \rangle $, $\chi _1=
1/\sqrt{2}(|\uparrow \downarrow \rangle +|\downarrow \uparrow
\rangle $, $\chi _2=|\downarrow \downarrow \rangle $,
$\chi_3=1/\sqrt{2}(|\ua \da \ra -|\da \ua \ra )$. We propose the
following quantum cloning transformations
\begin{eqnarray}
U\chi _0\otimes R =\sqrt {\frac {3}{4}}|3\uparrow \rangle \otimes
R_{\uparrow } +\sqrt {1\over 4}|2\uparrow ,\downarrow \rangle
\otimes R_{\downarrow},
\nonumber \\
U\chi _1\otimes R =\sqrt {\frac {1}{2}}|2\uparrow ,\downarrow
\rangle \otimes R_{\uparrow } +\sqrt {1\over 2}|\uparrow
,2\downarrow \rangle \otimes R_{\downarrow},
\nonumber \\
U\chi _2\otimes R =\sqrt {\frac {1}{4}}|\uparrow ,2\downarrow
\rangle \otimes R_{\uparrow } +\sqrt {3\over 4}|3\downarrow \rangle
\otimes R_{\downarrow},
\nonumber \\
U\chi _3\otimes R =\sqrt {\frac {1}{2}}|\widetilde {2\uparrow
,\downarrow }\rangle \otimes R_{\uparrow } +\sqrt {1\over
2}|\widetilde {\uparrow ,2\downarrow }\rangle \otimes
R_{\downarrow}, \label{clone}
\end{eqnarray}
where $R$s in the r.h.s. are ancillary and blank states, $|2\ua ,\da
\ra =(|\ua \ua \da \ra +|\ua \da \ua \ra + |\da \ua \ua \ra )/\sqrt
{3} $ is a symmetric state with 2 spins up and 1 spin down,
similarly for $|\ua ,2\da \ra $. The state $|\widetilde {2\ua ,\da
}\ra =(|\ua \ua \da \ra +\omega |\ua \da \ua \ra + \omega ^2|\da \ua
\ua \ra )/\sqrt {3}$ is almost the same as the symmetric state
$|2\ua ,\da \ra $ but with the phase of $\omega =e^{2\pi i/3}$.
$R_{\ua },R_{\da }$ are ancillary states and are orthogonal to each
other. It can be checked easily that the above relations satisfy the
unitary condition. We next show that this quantum cloning machine is
universal and optimal in the sense the relation (\ref{scalar}) is
satisfied and the shrinking factor saturates the optimal bound. We
expand the input state $\rho \otimes \rho $ in terms of the 4 basis
$\chi _i,i=0,1,2,3$. By using the cloning transformations
(\ref{clone}), tracing out the ancillary states $R_{\ua }, R_{\da
}$, we obtain the output state of 3 qubits. This state is a mixed
state and may be entangled. What we are interested is the reduced
density operator of each output qubit. One can see that each output
qubit is the same from the cloning transformation (\ref{clone}). By
some calculations (see the appendix for detail),  we
find the following relation,
\begin{eqnarray}
\rho ^{(out)}_{red.}=\frac {5}{6}\rho +\frac {1}{12}I.
\end{eqnarray}
Really, our cloning transformation (\ref{clone}) is universal and
optimal since the shrinking factor $\frac {5}{6}$ is optimal\cite{BEM}. This
is the first non-trivial quantum cloning of identical mixed qubits.
We remark that two identical pure qubits can be expanded in the
symmetric subspace, so the first three quantum cloning transformations
are enough for identical pure states. For general identical
mixed states, the cloning transformation for singlet state is
necessary.

{\it General 2 to $M$ $(M>2)$ UQCM.}-- Next, we shall present our
general result of 2 to $M$ cloning. The cloning machine creates $M$
copies out of 2 identical mixed qubits. The quantum cloning
transformation is presented as follows:
\begin{eqnarray}
U\chi _0\otimes R =\sum _{k=0}^{M-2}\alpha _{0k} |(M-k)\uparrow
,k\downarrow \rangle \otimes R_{k},
\nonumber \\
U\chi _1\otimes R =\sum _{k=0}^{M-2}\alpha _{1k} |(M-1-k)\ua
,(1+k)\da \rangle \otimes R_k, \nonumber
\\
U\chi _2\otimes R =\sum _{k=0}^{M-2}\alpha _{2k} |(M-2-k)\uparrow
,(2+k)\downarrow \rangle \otimes R_k \nonumber
\\
U\chi _3\otimes R =\sum _{k=0}^{M-2}\alpha _{1k} |\widetilde
{(M-1-k)\ua ,(1+k)\da }\rangle \otimes R_k, \label{gclone}
\end{eqnarray}
where
\begin{eqnarray}
\alpha _{jk}&=& \sqrt{\frac {6(M-2)!(M-j-k)!(j+k)!}{(2-j)!(M+1)!
(M-2-k)!j!k!}},\nonumber \\
&&~~~~~~~~~~~j=0, 1, 2. \label{parameter}
\end{eqnarray}
As previously, the state $|i\ua ,j\da \ra $ is a completely
symmetrical state with $i$ spins up and $j$ spins down, the state
$|\widetilde {i\ua ,j\da }\ra $ is almost the same as $|i\ua ,j\da
\ra $, but each term has a different phase of $\left(
\begin{array}{c}i+j \\ i\end{array}\right) $-th root of unity so
that $|i\ua ,j\da \ra $ and $|\widetilde {i\ua ,j\da }\ra $ are
orthogonal to each other. $R_k$ are ancillary states and are
orthogonal for different $k$. We can find that this quantum cloning
machine is universal and optimal, see appendix for detailed
calculations.
\begin{eqnarray}
\rho _{red.}^{(out)}= \frac {M+2}{2M}\rho +\frac {M-2}{4M}I,
\label{result}
\end{eqnarray}
where the shrinking factor $(M+2)/2M$ achieves the optimal
bound\cite{BEM}. Thus we show that we can copy two identical mixed
qubits as the same quality as we copy two identical pure states.

{\it Discussions and summary.}--In summary, we present the quantum
cloning transformations (\ref{gclone}) which can copy arbitrary two
identical mixed qubits. This quantum cloning machine is optimal
in the sense the shrinking factor between single input and
output qubit achieves the upper bound which is the same as for the
pure qubit.

The optimal quantum cloning is closely
related with quantum state estimation as presented in
Ref.\cite{BEM}. The optimal quantum state estimation are known for
identical pure states and the mixed state with support in symmetric
subspace. It is not clear how to make a state estimation for
identical mixed states which are not restricted to symmetric
subspace. In this paper, when $M\rightarrow \infty $, the quantum
cloning machine is naturally a realization of the quantum state
estimation. Since our cloning transformations work for arbitrary
identical mixed qubits (including identical pure states and mixed
state with support in symmetric subspace), we actually provide a
{\it universal} and {\it optimal} state estimation for this case.

{\it Acknowlegements:} One of the authors, H.F. was supported in part
by `Bairen' program and NSFC. He also would like to
thank V.Bu\v{z}ek and V. Roychowdhury for useful discussions and
encouragements.

{\it Appendix.}--First, we denote $A_{ij}=\chi _i\chi _j^{\dagger
}$. The density operator $\rho \otimes \rho $ can be written as,
\begin{eqnarray}
\rho \otimes \rho
&=&z_0^2A_{00}+z_1z_2\sqrt{2}A_{01}+z_1^2A_{02}\nonumber \\
&&+z_1z_2\sqrt{2}A_{10}+(z_0z_3+z_1z_2)A_{11}+z_1z_3\sqrt{2}A_{12}
\nonumber \\
&&+z_2^2A_{20}+z_2z_3\sqrt{2}A_{21}+z_3^2A_{22} \nonumber
\\&&+(z_0z_3-z_1z_2)A_{33}.
\label{expand}
\end{eqnarray}
To do quantum cloning for $\rho \times \rho $, we shall add blank
and ancillary state, do unitary transformation $U$ as presented in
Eqs.(\ref{clone},\ref{gclone}), then trace out the ancillary state.
The output state is written as
\begin{eqnarray}
\rho ^{(out)}=Tr_{R(k)}U(\rho \times \rho \otimes R)U^{\dagger },
\end{eqnarray}
where $Tr_{R(k)}$ means tracing out the ancillary state. Since the
cloning procedure is linear, we then can study the Eq.(\ref{expand})
term by term. We denote the output state of term $A_{ij}$ as $\rho
_{ij}$. Then the output state $\rho ^{(out)}$ is in the same form as
$\rho \otimes \rho $ in Eq.(\ref{expand}), the only difference is
that we should replace $A_{ij}$ by $\rho _{ij}$. By using the
cloning transformation (\ref{gclone}), we have
\begin{eqnarray}
\rho _{ij}&=&\sum _{k=0}^{M-2}\alpha _{ik}\alpha ^*_{jk}\left(
|(M-i-k)\ua ,(i+k)\da \rangle \right. \nonumber \\
&&\left. \langle (M-j-k)\ua ,(j+k)\da |\right)
, \nonumber \\
&&~~~~~~~i,j=0,1,2 \nonumber \\
\rho _{33}&=&\sum _{k=0}^{M-2}\alpha _{1k}\alpha ^*_{1k}\left(
|\widetilde {(M-1-k)\ua ,(1+k)\da }\rangle \right. \nonumber \\
&&\left. \langle \widetilde {(M-1-k)\ua ,(1+k)\da }|\right).
\end{eqnarray}
Thus by using the UQCM in Eq.(\ref{gclone}), we find explicitly the
output state $\rho ^{(out)}$.

Since we use the shrinking factor $f$ to quantify the quality of the
copies, we need to find the reduced density operator of single qubit
of the output state $Tr_{M-1}\rho ^{(out)}$. That means $M-1$ qubits
are traced out from the output state $\rho ^{(out)}$ and the single
qubit reduced density operator is obtained. We first consider the
diagonal elements of the reduced density operator. From the
definition of the symmetric state, we know that the state $|(M-i)\ua
,i\da \rangle $ can be rewritten as the following form,
\begin{eqnarray}
|(M-i)\ua ,i\da \rangle &=&\sqrt {\frac {C_{M-i}^i}{C_M^i}} |\ua
\rangle |(M-i-1)\ua ,i\da \rangle \nonumber \\
&+&\sqrt {\frac {C_{M-1}^{i-1}}{C_{N}^i}} |\da |(M-i)\ua ,(i-1)\da
\rangle \nonumber .
\end{eqnarray}
Since it is a symmetric state, each single qubit reduced density
operator is the same. It is written as
\begin{eqnarray}
&&Tr_{M-1}|(M-i)\ua ,i\da \rangle \ra \la (M-i)\ua ,i\da | \nonumber
\\
&=&\frac {C_{M-i}^i}{C_M^i}|\ua \ra \la \ua |+\frac
{C_{M-1}^{i-1}}{C_{N}^i} |\da \ra \la \da | \nonumber \\
&=&\frac {M-i}{M}|\ua \ra \la \ua |+\frac {i}{M} |\da \ra \la \da |.
\end{eqnarray}
With the help of the results in (\ref{parameter}), we know the
single qubit reduced density operator of $\rho _{ii}, i=0,1,2$ is
\begin{eqnarray}
&&Tr_{M-1}\rho _{ii}=\sum _{k=0}^{M-2}|\alpha _{ik}|^2 \left( \frac
{M-i-k}{M}|\ua \ra \la \ua |\right. \nonumber \\
&&~~~~~~~~~~~~~~~~~~~~~~~~~~~~\left. +\frac {i+k}{M}|\da \ra \la \da
|\right)
\nonumber \\
&=&\sum _{k=0}^{M-2}\frac {6(M-2)!}{(2-i)!i!(M+1)!}\frac
{(M-i-k)!(i+k)!}{(M-2-k)!k!} \times \nonumber \\
&&\times \left( \frac {M-i-k}{M}|\ua \ra \la \ua |+\frac
{i+k}{M}|\da \ra \la \da |\right).
\end{eqnarray}
Explicitly, we have the following results:
\begin{eqnarray}
Tr_{M-1}\rho _{00}&=&\frac {3M+2}{4M}|\ua \ra \la \ua | +\frac
{M-2}{4M}|\da \ra \la \da |, \nonumber \\
 Tr_{M-1}\rho _{11}&=&\frac
{1}{2}(|\ua \ra \la \ua | +|\da \ra \la \da |), \nonumber \\
Tr_{M-1}\rho _{22}&=&\frac {M-2}{4M}|\ua \ra \la \ua | +\frac
{3+2M}{4M}|\da \ra \la \da |.
\end{eqnarray}
The calculations for case $\rho _{33}$ are different from the case
$\rho _{11}$ since we have phases for each term in state
$|\widetilde { (M-1-k)\ua ,(1+k)\da }\rangle $. But by careful
analyzing, we find that these phases do not change the single qubit
reduced density operator, and we have
\begin{eqnarray}
Tr_{M-1}\rho _{33}=Tr_{M-1}\rho _{11}=\frac {1}{2}(|\ua \ra \la \ua
| +|\da \ra \la \da |).
\end{eqnarray}
Finally, let's study the off-diagonal elements of the reduced
density operator of $\rho ^{(out)}$. We have the following results:
\begin{eqnarray}
&&Tr_{M-1}\rho _{ii+1}=\sum _{k=0}^{M-2}\alpha _{ik}\alpha
^*_{i+1k}Tr_{M-1} |(M-i-k)\ua , \nonumber \\
&&(i+k)\da \rangle
\langle (M-i-1-k)\ua ,(i+1+k)\da |\nonumber \\
&=&\sum _{k=0}^{M-2}\alpha _{ik}\alpha ^*_{i+1k}\frac
{\sqrt {(M-i-k)(i+k+1)}}{M}|\ua \ra \la \da | \nonumber \\
&=&\frac {6}{M^2(M^2-1)}\frac {\sqrt {(2-i)(1+i)}}{(2-i)!(1+i)!}
\times \nonumber \\
&& \times \sum _{k=0}^{M-2}\frac {(M-i-k)!(i+k+1)!}{k!(M-2-i)!}|\ua
\ra \la \da |.
\end{eqnarray}
For cases $i=0,1$, we have
\begin{eqnarray}
Tr_{M-1}\rho _{01}=Tr_{M-1}\rho _{12}=\frac {\sqrt{2}(M+2)}{4M}|\ua
\ra \la \da |.
\end{eqnarray}
Similarly, we find
\begin{eqnarray}
Tr_{M-1}\rho _{10}=Tr_{M-1}\rho _{21}=\frac {\sqrt{2}(M+2)}{4M}|\da
\ra \la \ua |.
\end{eqnarray}
Summarize all of these results together, we have
\begin{eqnarray}
\rho _{red.}^{(out)}=Tr_{M-1}\rho ^{(out)}= \frac {M+2}{2M}\rho
+\frac {M-2}{4M}I.
\end{eqnarray}
This is the result presented in Eq.(\ref{result}).

\end{document}